\documentclass[preprint,showpacs,amsmath,amssymb,pra,superscriptaddress]{revtex4}
\usepackage{graphicx}% Include figure files
\usepackage{dcolumn}% Align table columns on decimal point
\usepackage{bm}% bold math
\begin{document}

%\preprint{APS/123-QED}

\title{Adiabatic radio frequency potentials for the coherent manipulation of matter waves}
 \author{I.~Lesanovsky}
\affiliation{Physikalisches Institut, Universit\"at Heidelberg,
D-69120 Heidelberg, Germany}
 \email{ilesanov@physi.uni-heidelberg.de}
  \author{T.~Schumm}
  \affiliation{Physikalisches Institut, Universit\"at Heidelberg, D-69120 Heidelberg, Germany}
  \affiliation{Laboratoire Charles Fabry de l'Institut d'Optique, UMR 8105 du CNRS, F-91403 Orsay, France}
 \author{S.~Hofferberth}
 \affiliation{Physikalisches Institut, Universit\"at Heidelberg, D-69120 Heidelberg, Germany}
  \author{L.~M.~Andersson}
 \affiliation{Physikalisches Institut, Universit\"at Heidelberg, D-69120 Heidelberg, Germany}
 \affiliation{Department of Microelectronics and Information  Technology,
 The Royal Institute of Technology, KTH, Electrum 229, SE-164 40, Kista, Sweden}
 \author{P.~Kr\"uger}
 \affiliation{Physikalisches Institut, Universit\"at Heidelberg, D-69120 Heidelberg, Germany}
 \affiliation{Laboratoire Kastler Brossel, 24 rue Lhomond, 75005 Paris, France}
 \author{J.~Schmiedmayer}
 \email{schmiedmayer@atomchip.org}
 \affiliation{Physikalisches Institut, Universit\"at Heidelberg, D-69120 Heidelberg, Germany}

\date{\today}

\begin{abstract}\label{txt:abstract}
Adiabatic dressed state potentials are created when magnetic
sub-states of trapped atoms are coupled by a radio frequency field.
We discuss their theoretical foundations and point out fundamental
advantages over potentials purely based on static fields. The
enhanced flexibility enables one to implement numerous novel
configurations, including double wells, Mach-Zehnder and Sagnac
interferometers which even allows for internal state-dependent atom
manipulation. These can be realized using simple and highly
integrated wire geometries on atom chips.
\end{abstract}

\maketitle
\section{Introduction}
Magnetic fields are powerful tools to control and manipulate the
motion of neutral atoms \cite{Migdall85,Wieman99}. These fields can
either be created by (macroscopic) coils \cite{Bergeman87}, free
standing wires \cite{Schmiedmayer95,Fortagh98,Denschlag99} or - as a
result of the growing effort for miniaturization and integration -
by surface-mounted micro fabricated structures, so-called atom chips
\cite{Folman02}. Compared to macroscopic setups, atom chips provide
high magnetic field gradients \cite{Reichel04} and therefore enable
the realization of tightly confining traps. The flexibility of
designing complex current and charge patterns on the chip allows for
considerable freedom to engineer `potential landscapes` for neutral
atoms. This has resulted in numerous designs of atom-optical
elements such as traps, guides, beams splitters and interferometers
\cite{Cassettari00,Muller00,Hinds01,Haensel01,Andersson02} with
possible applications ranging from quantum information processing
\cite{Calaro00,Cirone05} to high precision measurements
\cite{Kasevich91}. Even though many of these designs have been
demonstrated experimentally \cite{Folman02,Hommelhoff05,Guenther05},
there have been enormous difficulties to realize a coherent beam
splitter using microscopically tailored static or slowly varying
fields \cite{Shin05}.

Most of these difficulties stem from the fact that Maxwell's
equations constrain the freedom to design static magnetic
potentials. One consequence is that the number of potential minima
created is less or equal to the number of wires used \cite{Davis01}.
Whereas regular strongly confining potential minima are created from
quadrupole fields, the merging and splitting relies on higher order
multipoles, usually the hexapole component, and thus results in a
significantly weaker confinement. Consequently any dynamic splitting
of a potential passes through a weakly confining region and creates
an additional unwanted minimum, a loss channel. This splitting two
in two makes the central splitting region very unstable and
therefore truly adiabatic manipulations are hard to perform
\cite{Esteve05}.

These deficiencies can be overcome by using not only static fields
but combining them with oscillating radio frequency (RF) or
micro-wave near fields. The adiabatic dressed state potentials
created in this way do not show the unwanted loss channels, keep the
confinement tight during the splitting process and consequently
allow for a smooth transition from a single to two channels. Well
controlled coherent splitting and simultaneous tight confinement of
the atomic motion can be achieved even far away from the chip
surface \cite{Schumm05}. In addition adiabatic potentials permit the
creation of non-trivial topologies like, for example,
two-dimensional traps \cite{Zobay01,Coulombe04}, closed path
interferometers and ring geometries. Also smooth transformations
between different potential forms can be achieved.

In this paper we first discuss the theoretical foundations of the
underlying coupling creating the adiabatic potentials and present
their advantages.  These are then applied to create basic atom
optical elements such as a double-well, a Mach-Zehnder
interferometer and a ring trap. We also outline the implementation
of a state-dependent splitting scheme for atomic clouds.

\section{Theoretical description of dressed RF potentials}
We develop the theory by starting with the initial approach by Zobay
and Garraway \cite{Zobay01} and extending it to fully account for
the vector properties of the magnetic fields involved. Only
accounting for the latter leads to a complete description of the
underlying couplings and the increased versatility of the resulting
potentials.

We consider an alkali atom in a hyper-fine level designated by the
quantum number $F$. Assuming that $F$ remains a good quantum number
even in the presence of a magnetic field, the atomic dynamics is
governed by the Hamiltonian
\begin{eqnarray}
H_\text{initial}=g_F\mu_B\mathbf{B}(\mathbf{r},t)\cdot\mathbf{F}.\label{eq:initial_hamiltonian}
\end{eqnarray}
Here $g_F$ is the $g$-factor of the hyper-fine level and
$\mathbf{F}$ the angular momentum operator. We assume
$\mathbf{B}(\mathbf{r},t)$ to consist of a static part
$\mathbf{B}_S(\mathbf{r})$ and an oscillatory part of the form
\begin{eqnarray}
\mathbf{B}_O(\mathbf{r},t)=\mathbf{B}^A_\text{RF}(\mathbf{r})\cos(\omega
t)+\mathbf{B}^B_\text{RF}(\mathbf{r})\cos(\omega t+\gamma).
\end{eqnarray}
As a first step we use the unitary transformation $U_S$ to transform
the Hamiltonian into a frame where the interaction of the atom with
$\mathbf{B}_S(\mathbf{r})$ is diagonal, i.e.
\begin{eqnarray}
U^\dagger_S\mathbf{B}_S(\mathbf{r})\cdot\mathbf{F}U_S=\left[\mathfrak{R}_S\mathbf{B}_S(\mathbf{r})\right]\mathbf{F}=|\mathbf{B}_S(\mathbf{r})|F_z.
\end{eqnarray}
Here we have exploited that rotating the operator
$\mathbf{F}$ by using $U_S$ is equivalent to rotating the magnetic
field vector $\mathbf{B}_S(\mathbf{r})$ by applying the appropriate
rotation matrix $\mathfrak{R}_S$. The operator $F_z$ can be
represented as a diagonal matrix with the entries $-F \leq m_F \leq
F$ and $m_F$ denoting the magnetic hyper-fine sub-levels. We proceed
by applying another unitary operation
$U_R=\exp\left[-i\frac{g_F}{|g_F|}F_z \omega t\right]$ which
effectuates a transformation into a frame that rotates with the
angular velocity $\omega$ around the local direction of the static
field
$\mathbf{e}_S=\frac{\mathbf{B}_S(\mathbf{r})}{|\mathbf{B}_S(\mathbf{r})|}$.
The application of $U_R$ leads to the emergence of additional terms
that oscillate with the frequency $2\omega$. In the so-called
rotating wave approximation - which we employ in the following - the
oscillating terms are neglected. The now time-independent
Hamiltonian reads
\begin{eqnarray}
  H=\left[g_F\mu_B\left|\mathbf{B}_S(\mathbf{r})\right|-\frac{g_F}{|g_F|}\hbar\omega\right] F_z
  +\frac{g_F\mu_B}{2}\left(
     \begin{array}{c}
       \bar{B}_x \\
       \bar{B}_y \\
     \end{array}
   \right)^T
  \left(
     \begin{array}{c}
       F_x \\
       F_y \\
     \end{array}
   \right).\label{eq:hamiltonian_2RF}
\end{eqnarray}
The term proportional to $\hbar\omega F_z$ arises from the
transformation of the time derivative in the time-dependent
Schr\"odinger equation. The field $\bar{\mathbf{B}}$ is given by
\begin{eqnarray}
\bar{\mathbf{B}}=\mathfrak{R}_S\mathbf{B}^A_\text{RF}(\mathbf{r})
  +\mathfrak{R}_\delta\mathfrak{R}_S\mathbf{B}^B_\text{RF}(\mathbf{r})
\end{eqnarray}
where the matrix $\mathfrak{R}_\delta$ performs a rotation around
the axis $\mathbf{e}_S$ by the angle
\begin{eqnarray}
\delta=-\frac{g_F}{|g_F|}\gamma.
\end{eqnarray}
We want to emphasize that the sign of the rotation angle $\delta$
depends on the sign of the $g$-factor. Therefore atoms in different
hyperfine states will see different RF potentials even if they have
the same magnetic moment $\mu=m_F\times g_F$.

The adiabatic potentials are the eigenvalues of the Hamiltonian
(\ref{eq:hamiltonian_2RF}):
\begin{eqnarray}
 V_\text{ad}(\mathbf{r})=m_F^\prime\kappa\sqrt{\left[\left|\mathbf{B}_S(\mathbf{r})\right|-\frac{\hbar\omega}{|\kappa|}\right]^2
 +\frac{1}{4}\left[\bar{B}_x^2+\bar{B}_y^2\right]}\label{eq:adiabatic_potential}
\end{eqnarray}
with $\kappa=g_F\mu_B$.

In the case of zero phaseshift ($\gamma=0$), i.e. a linear polarized
RF field, the last term of the radical can be rewritten in a more
convenient form: $\bar{B}_x^2+\bar{B}_y^2=
\left[\mathbf{e}_S\times\left(\mathbf{B}^A_\text{RF}(\mathbf{r})
+\mathbf{B}^B_\text{RF}(\mathbf{r})\right)\right]^2$. Here it is
immediately apparent that only the RF field components being
perpendicular to the static field contribute.

\section{Realization of atom optical elements}
\subsection{Linear RF polarization - A double well}
\begin{figure}[htb]\center
\includegraphics[angle=0,width=8cm]{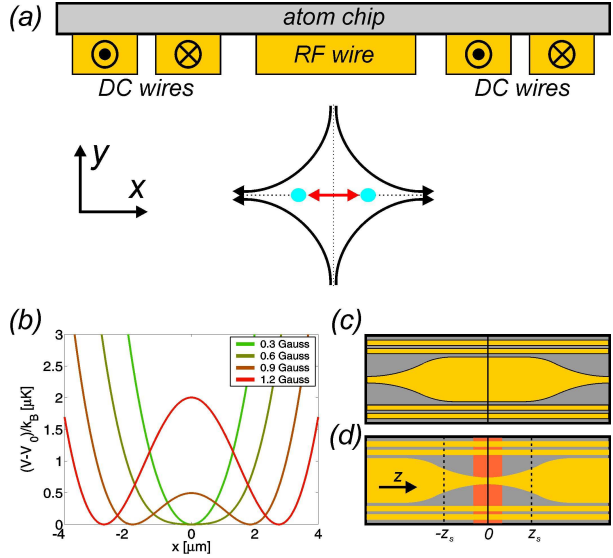}
\caption{(a) Experimental realization of the double-well
configuration. The quadrupole field is created by a surface-mounted
DC four-wire structure. The RF field is generated by a central broad
AC wire. Sufficiently close to the surface its RF field can
considered to be homogeneous. (b) Depending on the actual RF field
strength either a single- or a double-well is established. The
potential bottom of the individual curves has been subtracted. (c,d)
Longitudinally modulating the shape of the RF wire results in a
z-dependent variation of the RF amplitude. This can either be used
to achieve a confinement along the longitudinal (z-)axis (c) or a
spatially dependent splitting which would result in an
interferometer (d). Undesirable variations of the potential bottom
can be for example compensated by placing a charged wire underneath
the chip (red structure). }\label{fig:trap1}
\end{figure}

As a first example we consider the creation of a double-well
potential starting from a  Ioffe-Pritchard trap
\cite{Pritchard83,Folman02} which is one of the most commonly used
trapping field configuration. Its magnetic field is given by
\begin{eqnarray}
\mathbf{B}_S(\mathbf{r})=G\rho\left[\cos\phi\mathbf{e}_x-\sin\phi\mathbf{e}_y\right]+B_\text{I}\mathbf{e}_z.
\end{eqnarray}
Here $G$ is the gradient of the quadrupole field and $B_\text{I}$
the homogeneous Ioffe field strength. We superimpose a homogeneous
oscillatory RF field perpendicular to the Ioffe field. Without loss
of generality we take
$\mathbf{B}^A_\text{RF}(\mathbf{r})=B_\text{RF}\mathbf{e}_x$ and
$\mathbf{B}^B_\text{RF}(\mathbf{r})=0$. The unitary transformation
which diagonalizes the static part of the Hamiltonian
(\ref{eq:initial_hamiltonian}) is given by
\begin{eqnarray}
U_S=\exp\left[iF_z\phi\right]\exp\left[iF_y\beta\right]
\end{eqnarray}
with $\cos\beta=\frac{B_\text{I}}{|\mathbf{B}_S(\mathbf{r})|}$ and
$\sin\beta=-\frac{G\rho}{|\mathbf{B}_S(\mathbf{r})|}$ and
$|\mathbf{B}_S(\mathbf{r})|=\sqrt{G^2\rho^2+B_\text{I}^2}$. After
the transformation into the rotated frame the adiabatic potential
evaluates according to equation (\ref{eq:adiabatic_potential}) to
\begin{widetext}
\begin{eqnarray}
   V_\text{DW}(\mathbf{r})=m_F^\prime\kappa\sqrt{\left[\left|\mathbf{B}_S(\mathbf{r})\right|-\frac{\hbar\omega}{|\kappa|}\right]^2
   +\left[\frac{B_\text{RF}}{2|\mathbf{B}_S(\mathbf{r})|}\right]^2(B_\text{I}^2+G^2\rho^2\sin^2\phi)}.
\end{eqnarray}
\end{widetext}
Its minima are located at $\phi_1=0$ and $\phi_2=\pi$. Assuming that
$\rho\ll B_\text{I}/G$ \footnote{The validity of this condition can
be assured by applying a sufficiently large Ioffe field.} we can
approximate
\begin{eqnarray}
  V_\text{DW}(\rho,\phi_{1,2})=m_F^\prime\kappa\sqrt{\frac{G^4}{B_\text{I}^2}\left(\frac{\rho^2-\rho_0^2}{2}\right)^2+B_0^2}
\end{eqnarray}
with the position of the potential minimum
\begin{eqnarray}
\rho_0=\frac{1}{\sqrt{2}G}\sqrt{B_\text{RF}^2-B_\text{C}^2},
\end{eqnarray}
the potential bottom
\begin{eqnarray}
m_F^\prime\kappa
B_0=m_F^\prime\kappa\frac{B_\text{RF}}{4B_\text{I}}\sqrt{4
B_\text{I}^2+B_\text{C}^2-2G^2\rho_0^2}\approx m_F^\prime\kappa
\frac{B_\text{RF}}{2}\sqrt{1+\frac{\hbar\triangle}{|\kappa|
B_\text{I}}},\label{eq:potential_bottom}
\end{eqnarray}
the critical field strength
\begin{eqnarray}
B_\text{C}=2\sqrt{B_\text{I}\frac{\hbar\triangle}{|\kappa|}}
\end{eqnarray}
and the detuning
\begin{eqnarray}
\hbar\triangle=\kappa B_\text{I}-\hbar\omega.
\end{eqnarray}
In order to arrive at the last term of equation
(\ref{eq:potential_bottom}) we have exploited $G\rho_0\ll
B_\text{I}.$ For $B_\text{RF}\leq B_\text{C}$ the potential
$V_\text{DW}(\textbf{r})$ exhibits only a single well whereas for
$B_\text{RF}> B_\text{C}$ one encounters a double-well configuration
(see figure \ref{fig:trap1}b). The trap frequency in each well
evaluates approximately to
\begin{eqnarray}
\omega_\text{T,RF}=\sqrt{\frac{m_F^\prime\kappa}{m
B_0}}\frac{G^2\rho_0}{B_\text{I}}
\end{eqnarray}
with $m$ being the mass of the atom considered.

There are several advantages of a RF interferometer over a static
two-wire configuration \cite{Hinds01,Stickney03,Denschlag99_2}:
\begin{enumerate}
\item{The capability of performing a smooth transition from a true
single well to a double-well, by varying any of the parameters
$\triangle$, $B_\text{RF}$ and $B_\text{I}$. In contrast, in the
static case one encounters a transition from two vertically to two
horizontally split minima, if the strength of a homogeneous bias
field is modulated \cite{Hinds01}. In the vicinity of the splitting
region this leads to unwanted tunneling processes into the second
vertical (loss) channel just before the intended splitting sets in
\cite{Stickney03}. This poses a severe obstacle for establishing an
adiabatic process. In addition the RF realization of the double well
conserves the parabolic confinement perpendicular to the splitting
direction even in the vicinity of the splitting region. Here the
confinement in the static configuration in both directions relies
solely on a quartic potential.}

\item{In the static realization the distance between the potential
minima scales according to $\rho_0\propto \sqrt{b}$. Here $b$ is the
strength of a homogenous magnetic field which eventually gives rise
to the splitting of the hexapole into two quadrupole minima
\cite{Esteve05}. However, in order to have a stable splitting
distance one has to precisely control the field strength $b$, i.e.
keep its fluctuations $\triangle b$ small. This is extremely hard in
the vicinity of $b=0$ since $\triangle\rho_0/\triangle b\propto
b^{-1/2}$. Unlike this the splitting distance in the RF case obeys
$\rho_0\propto B_\text{RF}$ (for zero detuning). Thus we find
$\rho_0$ to be much less sensitive with respect to fluctuations in
$B_\text{RF}$ particularly if $B_\text{RF}$ is small.}

\item{The RF adiabatic potential keeps much tighter confining  wells
even far away from the field generating structures, i.e. the chip
surface. This can be illustrated considering an atom chip with
structure size $d$. For the sake of simplicity the quadrupole for
the RF setup shall be created by a sideguide configuration
\cite{Folman02} in a distance $d$ above the chip surface. The static
implementation of the double-well consists of two wires separated by
2d \cite{Esteve05}. Provided that the wire current $I$ and
$B_\text{I}$ are equal for both setups and assuming for simplicity
$\triangle=0$ the trap frequencies and the height of the barrier
between the wells obey
\begin{eqnarray}
  \frac{\omega_\text{T,RF}}{\omega_\text{T,static}}\propto\frac{d}{\rho_0}\sqrt{\frac{B_\text{RF}}{B_\text{I}}}\\
  \frac{h_\text{T,RF}}{h_\text{T,static}}\propto d^2\frac{G^2}{B_0B_I}.
  \label{eq:trap_frequency}
\end{eqnarray}
The essence of these expressions is their scaling with respect to
the parameter $d$ which refers not only to the structure size but
also to the distance of the potential wells to the chip surface.
Compared to the static two-wire case, a similar RF trap allows for
realizing the same confinement with larger structures and thus
farther away from the chip surface. The latter is of particular
importance as hereby coherence-destroying surface interactions
\cite{Henkel03,Schroll03} are strongly inhibited. The stronger
increase of the potential barrier in the RF case is advantageous as
it permits a true spatial separation of trapped atom clouds even for
small splitting distances}
\end{enumerate}
The potential of the RF technique to coherently control the motion
of atoms has recently enabled the demonstration of coherent
splitting of a Bose-Einstein Condensate (BEC) on an atom chip
\cite{Schumm05}.

In figure \ref{fig:trap1}a we present how a highly integrated
realization of a RF double-well could look like. The quadrupole
field is generated by a four-wire structure carrying
counter-propagating DC currents. In-between these wires there is a
broad wire flown through by an AC current. Sufficiently close to
this wire, the resultant RF field can be considered to be
homogeneous. The Ioffe field pointing into the plane of view is
generated by additional wires which are not shown here
\cite{Folman02}.
\begin{figure}[htb]\center
\includegraphics[angle=0,width=13cm]{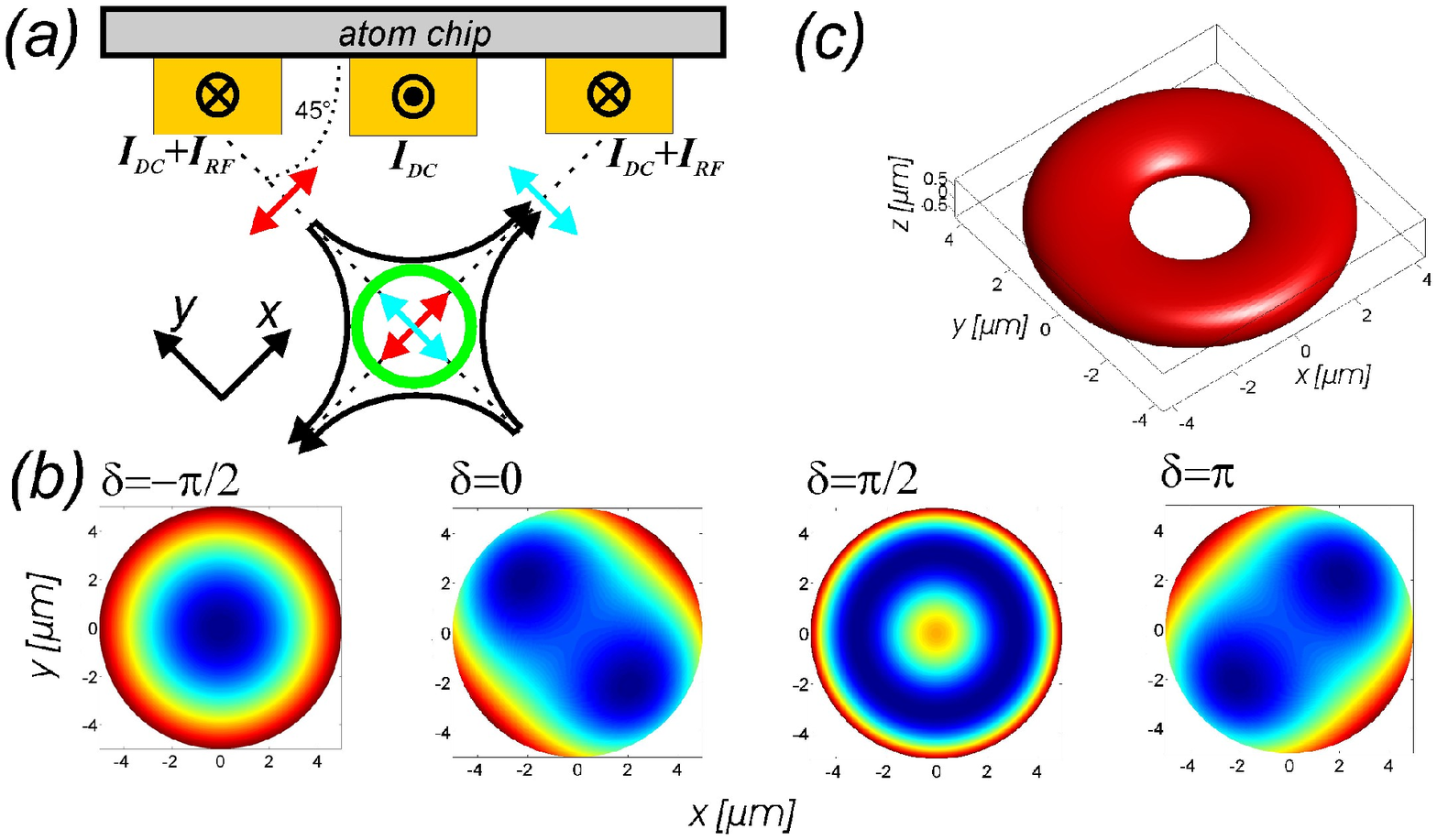}
\caption{(a) Experimental setup for realizing a ring shaped
potential. The static quadrupole field is generated by a three wire
configuration. The two outer wires also carry RF currents which
generate two phase shifted and orthogonally polarized oscillating
homogeneous fields in the vicinity of the quadrupole center. (b)
Depending on the phase shift $\delta$ either a single well, a
double-well or a ring-shaped potential emerge. (c) A 3D confinement
is achieved by introducing a spatially dependent RF amplitude of the
form
$B_\text{RF}(z)=1.3\,\text{Gauss}+0.05\,\frac{\text{Gauss}}{\text{m}^2}
z^2$. Visualization of the $m_Fg_FE=k_B\times 1.1\,\mu\text{K}$
isosurface for this case.}\label{fig:ring1}
\end{figure}
The potential bottom of the RF double-well increases proportional to
$(B_\text{RF}-B_\text{C})^2$. This provides an convenient mechanism
to achieve confinement in the longitudinal direction. A z-dependence
of the RF amplitude, i.e. $B_\text{RF}=B_\text{RF}(z)$, can be
achieved by shaping the RF wire \cite{Kraft05}. For example a wire
geometry creating a symmetric increase of the current density around
$z=0$, consequently, will lead to a symmetric increase of the RF
amplitude (see figure \ref{fig:trap1}c). Hence, depending on the
actual value of $B_\text{RF}$ a three-dimensionally confining
single- or double-well is achieved. Similarly a Mach-Zehnder
interferometer can be realized by varying the RF amplitude such that
$B_\text{RF}(0)>B_\text{C}$ and $B_\text{RF}(z)|_{|z|\geq
z_\text{S}}<B_\text{C}$ with $z_\text{S}$ defining the length of the
splitting region as indicated in figure \ref{fig:trap1}d. The
variations of the potential bottom can be compensated by applying
either a spatially varying Ioffe field or an additional external
potential. The latter can be realized for instance by placing a
charged wire underneath the chip \cite{Krueger03}. The corresponding
electric potential reads $U_\text{el}(\mathbf{r})=-\frac{\alpha}{2}
\left|\mathbf{E}(\mathbf{r})\right|^2$ \cite{Folman02}.

\subsection{Arbitrary RF polarization - A ring interferometer}
As a second example we construct a more complex trapping geometry by
employing two phase-shifted RF fields. We consider two orthogonal RF
fields of the form
$\mathbf{B}^A_\text{RF}(\mathbf{r})=\frac{B_\text{RF}}{\sqrt{2}}\mathbf{e}_x$
and
$\mathbf{B}^B_\text{RF}(\mathbf{r})=\frac{B_\text{RF}}{\sqrt{2}}\mathbf{e}_y$,
which are superimposed on the static $\mathbf{B}_S(\mathbf{r})$.
According to equation (\ref{eq:adiabatic_potential}) the
corresponding adiabatic potential evaluates to
\begin{widetext}
\begin{eqnarray}
  V_\text{R}(\mathbf{r})=m_F^\prime\kappa\sqrt{\left[\left|\mathbf{B}_S(\mathbf{r})\right|-\frac{\hbar\omega}{|\kappa|}\right]^2
  +\frac{B_\text{RF}^2}{8\left|\mathbf{B}_S(\mathbf{r})\right|^2}
  \left[G^2\rho^2(1+\sin(2\phi)\cos\delta)+2B_\text{I}(B_\text{I}+\left|\mathbf{B}_S(\mathbf{r})\right|\sin\delta)\right]}.\label{eq:ring_dw}
\end{eqnarray}
\end{widetext} For $\cos\delta>0$ we find the minima and maxima at
of the potential at
$\phi_\text{min}=\frac{3}{4}\pi,\frac{7}{4}\pi$ and
$\phi_\text{max}=\frac{1}{4}\pi,\frac{5}{4}\pi$, respectively. If
$\cos\delta<0$ the positions of the minima and maxima simply
exchange. Assuming $\rho\ll B_\text{I}/G$ the radial position of
these extrema is
\begin{eqnarray}
\rho_0=\frac{1}{2G}\sqrt{B_\text{RF}^2(1-\cos\delta\sin(2\phi)+\sin\delta)-2B^2_\text{C}}.
\end{eqnarray}
Hence for $\cos\delta>0$ and
$B_\text{RF}<\sqrt{\frac{2}{1+\cos\delta+\sin\delta}}B_\text{C}$
or $\cos\delta<0$ and
$B_\text{RF}<\sqrt{\frac{2}{1-\cos\delta+\sin\delta}}B_\text{C}$
solely a single minimum can be achieved. For
$\delta=\frac{3}{2}\pi$ in any case only a single minimum is
found.

In figure \ref{fig:ring1}a we present how such setup can be realized
in a highly integrated way. The static quadrupole field is generated
by a three wire setup. The two outer wires also serve as RF sources
that are positioned such that two orthogonally polarized homogeneous
fields in the vicinity of the quadrupole are created.

The versatility of the potential (\ref{eq:ring_dw}) lies in the fact
that by simply varying the phase shift $\delta$, i.e. changing the
polarization of the RF field, one can either accomplish a single
well, a double-well or a ring configuration. Even a rotating
double-well is achievable by appropriately tuning the phase and the
RF amplitude. The double-well configuration with the strongest
confinement is achieved for $\gamma=0$, i.e. vanishing relative
phase shift of the RF fields. Increasing the phase shift from
$\delta=0$ to $\delta=\frac{\pi}{2}$, i.e. from linear to circular
polarization, results in a smooth transition to a ring-shaped
potential of adjustable radius. This transition is shown in figure
\ref{fig:ring1}b. The potentials shown are calculated for the
typical set of experimental parameters $B_\text{I}=1\,
\text{Gauss}$, $G=0.2\, \text{Gauss}/\mu\text{m}$,
$B_\text{RF}=1.3\, \text{Gauss}$ and $\omega=2\pi\times
1.26\,\text{MHz}\,$. In order to generate a confinement also in the
longitudinal (z-)direction we impose a modulation of the RF
amplitude of the form
$B_\text{RF}(z)=\left(B_\text{RF}+G^2_\text{RF} z^2\right)$. In
figure \ref{fig:ring1}c the $m_Fg_FE=k_B\times 1.1\,\mu\text{K}$
isosurface for $G^2_\text{RF}=0.05\, \text{Gauss}/\text{m}^2$ is
depicted. The ring-shaped potential is thus capable of confining
BECs as the typical energy scale associated to such matter waves is
in the $\text{nK}$-regime.

The setup allows one to examine the collective ground state of
ultra-cold atoms trapped on a ring \cite{Alon04}. Also building a
ring interferometer (Sagnac-interferometer) for matter waves is
possible. Coherence preserving loading of the latter could be done
by preparing an ultra-cold atomic ensemble in the static wire trap.
Switching on the RF fields thereafter, and establishing the
appropriate phase shift $\delta$ leads to a well controlled
transition to the ring-shaped potential. Such traps are particularly
suited for building gyroscopes or rotation-sensors. Gupta \textit{et
al.} \cite{Gupta05} have recently succeeded in loading a ring-shaped
waveguide with a BEC. Their setup consists of millimeter-sized coils
forming a magnetic quadrupole ring with diameters ranging from 1.2
to 3 mm. However, generating BECs which are phase coherent over the
entire ring is extremely difficult in such a macroscopic trap. In
order to avoid the necessity of cooling to extremely low
temperatures it is beneficial to use small rings with diameters of a
few micrometers.

Also internal state-dependent manipulation of atoms can be achieved
by using the potential (\ref{eq:ring_dw}). Let us consider for
instance the two hyperfine states $\left|1\right>$ and
$\left|2\right>$ with the same magnetic moment $\mu = m_F g_F
\mu_B$. Consequently in a static field atoms in either of these
states are subjected to the same trapping potential. Suppose now the
RF field is switched on adiabatically such that $m_F^\prime=m_F$. In
the case when ${g_F}_{\left|1\right>} = - {g_F}_{\left|2\right>}$ we
have $\delta_{\left|1\right>}=-\delta_{\left|2\right>}$. Thus if one
picks $\delta_{\left|1\right>}=\frac{\pi}{2}$ atoms being in state
$\left|1\right>$ see a ring whereas atoms in state $\left|2\right>$
are confined to a single centered potential minimum as seen in
figure \ref{fig:ring1}b.

\section{Conclusion}
In conclusion dressed RF adiabatic potentials are versatilely
applicable to build atom optical elements and offer a number of
significant advantages over their static implementations. RF-based
traps provide tight confinement even at large surface distances and
allow for a smooth transition from a single to a double-well.
Moreover, a RF double-well is more robust against experimental
fluctuations against its static counterpart which is certainly
advantageous for performing tunneling experiments. This technique
paves the way to the realization of complex coherence preserving
potentials on a micro scale by using simple and highly integrated
setups. This is of particular importance for such demanding
applications as quantum information processing and high precision
measurements based on matter wave interference.

After submission of this manuscript several other works appeared
which discuss applications of RF potential
\cite{Fernholz05,Courteille05,Morizot05}.

We acknowledge financial support from the European Union, contract
numbers IST-2001-38863 (ACQP), MRTN-CT-2003-505032 (Atom Chips),
HPMF-CT-2002-02022, and the Deutsche Forschungsgemeinschaft,
contract number SCHM 1599/1-1. P.K. acknowledges support from the
Alexander von Humboldt foundation.

\end{document}